\documentclass[preprint,3p]{elsarticle}

\usepackage{epsfig}
\usepackage{axodraw}
\usepackage{amsmath}
\usepackage{amssymb}

\def\Cpl#1#2{\, #1 \! \cdot \! #2 \,}
\def\CplPow#1#2#3{\Cpl{#1}{#2}\!\!{}^#3}

\begin{document}

\begin{frontmatter}


\title{Precision calculation of processes used for luminosity measurement at the ZEUS experiment}

\author[TH]{T. Haas\fnref{fnTH}}
\address[TH]{DESY Hamburg}
\author[VM]{V. Makarenko\corref{corVM}\fnref{fnVM}}
\address[VM]{NCPHEP Minsk}
\fntext[fnTH]{now at XFEL}
\fntext[fnVM]{supported by DESY}
\cortext[corVM]{E-mail:Vladimir.Makarenko@desy.de}

\begin{abstract}
The process $p e^{\pm} \to p e^{\pm} \gamma$ with the photon emitted along the electron beam axis
is used for luminosity measurement at HERA.
In this paper the process is calculated including
one-loop QED radiative corrections.
In the ZEUS experiment, both the electron and the photon can be detected.
Therefore both photon and electron spectra with and without the $\gamma-e$ coincidence are analyzed.
We also calculate the process $p e^{\pm} \to p e^{\pm} \: l^- l^+$
which contributes to the background in the electron tagger.
\end{abstract}


\end{frontmatter}

\section{Introduction}

The luminosity in the ZEUS experiment at the HERA electron-positron collider
is measured using the inclusive bremsstrahlung process $p e^{\pm} \to p e^{\pm} \gamma$
at very small photon angles ($\Theta_{\gamma} \sim 0.1$ mrad).
A photon spectrometer system is located on the electron beam axis at a distance of about $100$ m from the interaction point~\cite{ZeusLumi}.
Additional detectors are used to measure forward-going
electrons\footnote{In this paper we use the generic term "electron" for any final state lepton with charge identical to the initial state lepton.}
with energies between $4-9$~GeV.
The electron rate is used to calibrate the acceptance of the photon spectrometer.
This study is motivated by augmented analysis precision which makes it necessary to go beyond the classical Bethe-Heitler formula~\cite{BHOrig}
used for simulation in ZEUS so far.
These simulations assume scattering on a point-like, spin-less proton.
No proton recoil and no higher-order effects were taken into account.

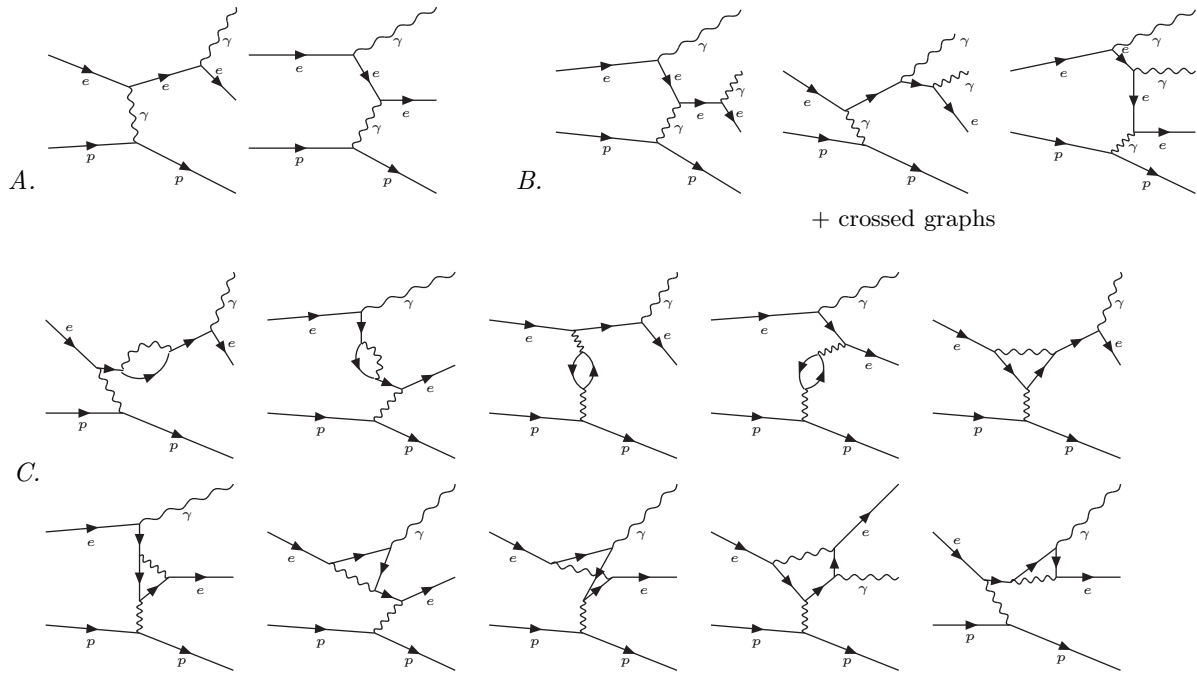
\begin{figure}[tb]
\centering
\begin{picture}(450,250)
\SetOffset(15,180) \ArrowLine(0,17)(33,19)\Text(16,13)[c]{\tiny $p$}
\ArrowLine(0,52)(30,40)\Text(14,42)[c]{\tiny $e$} \ArrowLine(33,19)(70,0)\Text(49,5)[c]{\tiny $p$}
\ArrowLine(57,48)(70,35)\Text(64,46)[c]{\tiny $e$} \Photon(57,48)(70,70){1}{4}\Text(65,57)[l]{\tiny $\gamma$}
\ArrowLine(30,40)(57,48)\Text(44,40)[c]{\tiny $e$} \Photon(33,19)(30,40){1}{4}\Text(34,29)[l]{\tiny $\gamma$}
\SetOffset(90,180) \ArrowLine(0,17)(39,17)\Text(19,12)[c]{\tiny $p$}
\ArrowLine(0,52)(39,52)\Text(19,47)[c]{\tiny $e$} \ArrowLine(39,17)(70,0)\Text(52,4)[c]{\tiny $p$}
\ArrowLine(49,35)(70,35)\Text(59,30)[c]{\tiny $e$} \Photon(39,52)(70,70){1}{4}\Text(56,57)[c]{\tiny $\gamma$}
\Photon(49,35)(39,17){1}{4}\Text(46,24)[l]{\tiny $\gamma$} \ArrowLine(39,52)(49,35)\Text(46,45)[l]{\tiny $e$}
\SetOffset(15,180) \Text(-15,5)[l]{\it A.}
\SetOffset(205,180) \ArrowLine(0,23)(38,19)\Text(19,16)[c]{\tiny $p$}
\ArrowLine(0,46)(38,50)\Text(19,43)[c]{\tiny $e$} \ArrowLine(38,19)(69,0)\Text(51,5)[c]{\tiny $p$}
\ArrowLine(62,34)(69,23)\Text(67,30)[l]{\tiny $e$} \Photon(62,34)(69,46){1}{4}\Text(67,38)[l]{\tiny $\gamma$}
\Photon(38,50)(69,69){1}{4}\Text(55,55)[c]{\tiny $\gamma$} \ArrowLine(46,34)(62,34)\Text(54,29)[c]{\tiny $e$}
\Photon(38,19)(46,34){1}{4}\Text(44,24)[l]{\tiny $\gamma$} \ArrowLine(38,50)(46,34)\Text(44,44)[l]{\tiny $e$}
\SetOffset(290,180) \ArrowLine(0,23)(31,18)\Text(15,15)[c]{\tiny $p$} \ArrowLine(0,46)(23,31)\Text(9,34)[c]{\tiny $e$}
\ArrowLine(31,18)(69,0)\Text(48,5)[c]{\tiny $p$} \ArrowLine(56,40)(69,23)\Text(70,27)[l]{\tiny $e$}
\Photon(56,40)(69,46){1}{4}\Text(70,40)[c]{\tiny $\gamma$} \Photon(44,42)(64,60){1}{4}\Text(66,57)[l]{\tiny $\gamma$}
\ArrowLine(44,42)(56,40) \Photon(31,18)(23,31){1}{4}\Text(29,26)[l]{\tiny $\gamma$} \ArrowLine(23,31)(44,42)
\SetOffset(375,180) \ArrowLine(0,23)(38,15)\Text(18,14)[c]{\tiny $p$}
\ArrowLine(0,46)(38,54)\Text(20,45)[c]{\tiny $e$} \ArrowLine(38,15)(69,0)\Text(51,3)[c]{\tiny $p$}
\ArrowLine(46,23)(69,23)\Text(57,18)[c]{\tiny $e$} \Photon(46,46)(69,46){1}{4}\Text(57,41)[c]{\tiny $\gamma$}
\Photon(38,54)(69,69){1}{4}\Text(55,57)[c]{\tiny $\gamma$} \Photon(46,23)(38,15){1}{4}\Text(46,17)[c]{\tiny $\gamma$}
\ArrowLine(46,46)(46,23)\Text(49,36)[l]{\tiny $e$} \ArrowLine(38,54)(46,46)\Text(43,54)[c]{\tiny $e$}
\SetOffset(205,180) \Text(-15,5)[l]{\it B.}
\SetOffset(290,180) \Text(45,-10)[c]{\small + crossed graphs}
\SetOffset(14,80) \Text(-12,-5)[l]{\it C.}
\SetOffset(14,80) \ArrowLine(0,17)(29,17)\Text(14,12)[c]{\tiny $p$}
\ArrowLine(0,52)(19,34)\Text(9,49)[c]{\tiny $e$} \ArrowLine(29,17)(70,0)\Text(48,4)[c]{\tiny $p$}
\ArrowLine(62,48)(70,35)\Text(68,43)[l]{\tiny $e$} \Photon(62,48)(70,70){1}{4}\Text(68,58)[l]{\tiny $\gamma$}
\ArrowLine(46,40)(62,48) \Photon(29,17)(19,34){1}{4} \ArrowLine(19,34)(29,33) \PhotonArc(40,32)(11,52,172){1}{4.5} \ArrowArc(35,41)(11,232,352)
\SetOffset(97,80) \ArrowLine(0,17)(40,14)\Text(20,10)[c]{\tiny $p$}
\ArrowLine(0,52)(35,55)\Text(17,48)[c]{\tiny $e$} \ArrowLine(40,14)(70,0)\Text(53,3)[c]{\tiny $p$}
\ArrowLine(50,26)(70,35)\Text(61,26)[c]{\tiny $e$} \Photon(35,55)(70,70){1}{4}\Text(53,58)[c]{\tiny $\gamma$}
\Photon(50,26)(40,14){1}{4} \ArrowLine(40,30)(50,26) \ArrowLine(35,55)(35,43) \PhotonArc(34,35)(8,-39,81){1}{4.5} \ArrowArc(41,38)(8,141,261)
\SetOffset(180,80) \ArrowLine(0,17)(35,14)\Text(17,10)[c]{\tiny $p$}
\ArrowLine(0,52)(31,48)\Text(15,45)[c]{\tiny $e$} \ArrowLine(35,14)(70,0)\Text(51,3)[c]{\tiny $p$}
\ArrowLine(57,51)(70,35)\Text(65,46)[l]{\tiny $e$} \Photon(57,51)(70,70){1}{4}\Text(65,58)[l]{\tiny $\gamma$}
\ArrowLine(31,48)(57,51) \Photon(35,14)(35,26){1}{4} \Photon(31,48)(35,40){1}{4} \ArrowArc(31,33)(8,300,420) \ArrowArc(39,33)(8,120,240)
\SetOffset(263,80) \ArrowLine(0,17)(35,14)\Text(17,10)[c]{\tiny $p$}
\ArrowLine(0,52)(40,55)\Text(20,48)[c]{\tiny $e$} \ArrowLine(35,14)(70,0)\Text(51,3)[c]{\tiny $p$}
\ArrowLine(50,43)(70,35)\Text(59,35)[c]{\tiny $e$} \Photon(40,55)(70,70){1}{4}\Text(57,58)[c]{\tiny $\gamma$}
\ArrowLine(40,55)(50,43) \Photon(50,43)(40,39){1}{4} \Photon(35,14)(35,26){1}{4} \ArrowArc(41,31)(8,99,219) \ArrowArc(34,34)(8,279,399)
\SetOffset(346,80) \ArrowLine(0,17)(35,14)\Text(17,10)[c]{\tiny $p$}
\ArrowLine(0,52)(23,40)\Text(9,42)[c]{\tiny $e$} \ArrowLine(35,14)(70,0)\Text(51,3)[c]{\tiny $p$}
\ArrowLine(62,48)(70,35)\Text(68,43)[l]{\tiny $e$} \Photon(62,48)(70,70){1}{4}\Text(68,58)[l]{\tiny $\gamma$}
\ArrowLine(46,40)(62,48) \Photon(35,14)(35,26){1}{4} \Photon(23,40)(46,40){1}{4} \ArrowLine(23,40)(35,26) \ArrowLine(35,26)(46,40)
\SetOffset(14,0) \ArrowLine(0,17)(35,14)\Text(17,10)[c]{\tiny $p$}
\ArrowLine(0,52)(35,55)\Text(17,48)[c]{\tiny $e$} \ArrowLine(35,14)(70,0)\Text(51,3)[c]{\tiny $p$}
\ArrowLine(46,35)(70,35)\Text(58,30)[c]{\tiny $e$} \Photon(35,55)(70,70){1}{4}\Text(53,58)[c]{\tiny $\gamma$}
\ArrowLine(35,26)(46,35) \Photon(46,35)(35,43){1}{4} \Photon(35,14)(35,26){1}{4} \ArrowLine(35,55)(35,43) \ArrowLine(35,43)(35,26)
\SetOffset(97,0) \ArrowLine(0,17)(40,14)\Text(20,10)[c]{\tiny $p$}
\ArrowLine(0,52)(23,40)\Text(9,42)[c]{\tiny $e$} \ArrowLine(40,14)(70,0)\Text(53,3)[c]{\tiny $p$}
\ArrowLine(50,26)(70,35)\Text(61,26)[c]{\tiny $e$} \Photon(46,46)(70,70){1}{4}\Text(57,51)[c]{\tiny $\gamma$}
\Photon(50,26)(40,14){1}{4} \ArrowLine(40,30)(50,26) \ArrowLine(23,40)(46,46) \ArrowLine(46,46)(40,30) \Photon(23,40)(40,30){1}{4}
\SetOffset(180,0) \ArrowLine(0,17)(35,14)\Text(17,10)[c]{\tiny $p$}
\ArrowLine(0,52)(23,40)\Text(9,42)[c]{\tiny $e$} \ArrowLine(35,14)(70,0)\Text(51,3)[c]{\tiny $p$}
\ArrowLine(46,35)(70,35)\Text(58,30)[c]{\tiny $e$} \Photon(46,46)(70,70){1}{4}\Text(57,51)[c]{\tiny $\gamma$}
\Photon(46,35)(23,40){1}{4} \ArrowLine(35,26)(46,35) \Photon(35,14)(35,26){1}{4} \ArrowLine(23,40)(46,46) \ArrowLine(46,46)(35,26)
\SetOffset(263,0) \ArrowLine(0,17)(35,14)\Text(17,10)[c]{\tiny $p$}
\ArrowLine(0,52)(23,40)\Text(9,42)[c]{\tiny $e$} \ArrowLine(35,14)(70,0)\Text(51,3)[c]{\tiny $p$}
\ArrowLine(46,46)(70,70)\Text(58,52)[c]{\tiny $e$} \Photon(46,35)(70,35){1}{4}\Text(57,30)[c]{\tiny $\gamma$}
\ArrowLine(46,35)(46,46) \Photon(46,46)(23,40){1}{4} \Photon(35,14)(35,26){1}{4} \ArrowLine(35,26)(46,35) \ArrowLine(23,40)(35,26)
\SetOffset(346,0) \ArrowLine(0,17)(29,17)\Text(14,12)[c]{\tiny $p$}
\ArrowLine(0,52)(19,34)\Text(9,49)[c]{\tiny $e$} \ArrowLine(29,17)(70,0)\Text(48,4)[c]{\tiny $p$}
\ArrowLine(46,35)(70,35)\Text(58,30)[c]{\tiny $e$} \Photon(46,46)(70,70){1}{4}\Text(57,51)[c]{\tiny $\gamma$}
\ArrowLine(46,46)(46,35) \Photon(46,35)(29,33){1}{4} \Photon(29,17)(19,34){1}{4} \ArrowLine(29,33)(46,46) \ArrowLine(19,34)(29,33)
\end{picture}
\caption{
The lowest order diagrams for $p e^{\pm} \to p e^{\pm} \gamma$~(A)
and $p e^{\pm} \to p e^{\pm} \gamma \gamma$~(B) processes
and QED loop diagrams considered~(C).
}
\label{fig_Diags}
\end{figure}

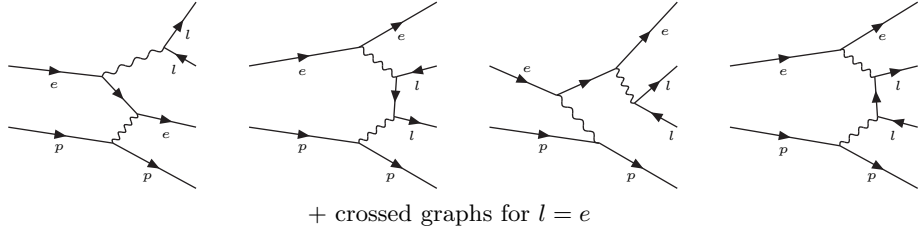
\begin{figure}[thb]
\centering
\begin{picture}(450,80)
\SetOffset(60,10) \ArrowLine(0,23)(39,17)\Text(19,15)[c]{\tiny $p$} \ArrowLine(0,46)(35,42)\Text(17,39)[c]{\tiny $e$}
\ArrowLine(39,17)(70,0)\Text(52,4)[c]{\tiny $p$} \ArrowLine(48,28)(70,23)\Text(59,20)[c]{\tiny $e$}
\ArrowLine(70,46)(58,53)\Text(62,45)[c]{\tiny $l$} \ArrowLine(58,53)(70,70)\Text(66,58)[l]{\tiny $l$}
\Photon(48,28)(39,17){1}{4} \ArrowLine(35,42)(48,28) \Photon(58,53)(35,42){1}{4}
\SetOffset(150,10) \ArrowLine(0,23)(41,17)\Text(20,15)[c]{\tiny $p$} \ArrowLine(0,46)(41,53)\Text(20,44)[c]{\tiny $e$}
\ArrowLine(41,17)(70,0)\Text(53,4)[c]{\tiny $p$} \ArrowLine(54,27)(70,23)\Text(61,20)[c]{\tiny $l$}
\ArrowLine(70,46)(55,42)\Text(63,39)[c]{\tiny $l$} \ArrowLine(41,53)(70,70)\Text(57,57)[c]{\tiny $e$}
\Photon(54,27)(41,17){1}{4} \ArrowLine(55,42)(54,27) \Photon(41,53)(55,42){1}{4}
\SetOffset(240,10) \ArrowLine(0,23)(41,17)\Text(20,15)[c]{\tiny $p$} \ArrowLine(0,46)(25,35)\Text(14,45)[c]{\tiny $e$}
\ArrowLine(41,17)(70,0)\Text(53,4)[c]{\tiny $p$} \ArrowLine(47,45)(70,70)\Text(64,59)[l]{\tiny $e$}
\ArrowLine(70,23)(54,32)\Text(67,20)[l]{\tiny $l$} \ArrowLine(54,32)(70,46)\Text(67,39)[c]{\tiny $l$}
\Photon(47,45)(54,32){1}{4} \ArrowLine(25,35)(47,45) \Photon(41,17)(25,35){1}{4}
\SetOffset(330,10) \ArrowLine(0,23)(41,16)\Text(20,14)[c]{\tiny $p$} \ArrowLine(0,46)(41,52)\Text(20,44)[c]{\tiny $e$}
\ArrowLine(41,16)(70,0)\Text(53,4)[c]{\tiny $p$} \ArrowLine(54,42)(70,46)\Text(63,39)[c]{\tiny $l$}
\ArrowLine(70,23)(55,27)\Text(61,20)[c]{\tiny $l$} \ArrowLine(41,52)(70,70)\Text(57,57)[c]{\tiny $e$}
\Photon(54,42)(41,52){1}{4} \ArrowLine(55,27)(54,42) \Photon(41,16)(55,27){1}{4}
\SetOffset(225,0) \Text(0,0)[c]{\small + crossed graphs for $l=e$}
\end{picture}
\caption{ The diagrams for $p e^{\pm} \to p e^{\pm} l^{+} l^{-}$ process. } \label{fig_Diags_Peee}
\end{figure}

In this paper we make an analysis of the processes affecting both the photon detector and the electron tagger in ZEUS.
Two experimental configurations are considered, with and without coincidence of electron and photon.
We consider the process $p e^{\pm} \to p e^{\pm} \gamma$ with the photon emitted along the electron-beam axis.
The one-loop QED radiative correction (Fig.~\ref{fig_Diags}) to the leptonic current is calculated.
The reaction in which a lepton pair is produced, $p e^{\pm} \to p e^{\pm} l^+ l^-$ (Fig.~\ref{fig_Diags_Peee})
with at least one detectable electron, is also considered.
It does not affect the photon spectrometer but is required for the proper simulation of the electron tagger signals.
Finally we discuss the systematic error of the luminosity measurement originating from the uncertainty in the cross section calculation.

In previous theoretical estimates for these processes~\cite{BH_Gaemers88,BH_Horst90}
good agreement with the classical formula was observed.
The contribution of higher-order effects appeared within the $1\%$-error region.
However, these calculations were made for different experimental configurations (coincidence between electron and photon)
and for a narrow photon energy range ($8 < E_{\gamma} < 14$~GeV).
No generator program was published that can be used for detector simulations.

This study is restricted to the QED leptonic current corrections only.
Weak effects are suppressed by the $Z$-boson mass in the considered momentum transfer region ($Q^2 < 10^{-5} \text{GeV}^2$).
The two-photon exchange vanishes in the forward region according to the existing estimate~\cite{Blunden03}.
The photons radiated by the initial or final state proton are directed along the proton beam axis and escape the calorimeter.
The other processes which are discussed in~\cite{BH_Gaemers89,BH_Horst90b} may also be neglected.

All results in this paper are compared to the classical Bethe-Heitler formula~\cite{BHOrig}:
\begin{equation}\label{classic_E}
\frac{d \sigma}{d E_{\gamma}} = 4 \alpha r_e^2 \frac{E_e'}{E_{\gamma} E_e}
\left(\frac{E_e}{E_e'}+\frac{E_e'}{E_e}-\frac{2}{3}\right)
\left(\ln{\frac{4 E_p E_e E_e'}{m_p m_e E_{\gamma}}}-\frac{1}{2}\right).
\end{equation}
Here $E_e$($E_e'$) is the energy of the initial(final) electron,
$\alpha$ is the fine structure constant and $r_e = \alpha /m_e$ is the classical electron radius.
The approximation for the angular distribution~\cite{BH_GandH} is also checked:
\begin{equation}\label{classic_Theta}
\frac{d \sigma}{d \Theta_{\gamma}} \sim \frac{\Theta_{\gamma}}{\left( (m_e/E_e)^2 + \Theta_{\gamma}^2 \right)^2}.
\end{equation}

The paper is organized as follows:
The kinematics are discussed in Sec.~\ref{sec_Kinematics}
where it is shown that kinematic variables may differ by up to a factor $10^{30}$ at opposite edges of the phase space.
The matrix element calculation is discussed in Sec.~\ref{sec_ME}.
Using those results a generator program has been created~\cite{hbgen_program}
which is described in Sec.~\ref{sec_Generator}.
This section also describes the precision-saving algorithm for the phase space generation procedure.
This algorithm is necessary since the computation is complicated by large numerical cancellations at small scattering angles.
The results are shown in Sec.~\ref{sec_Results}.
The paper closes with a conclusion.

\section{Kinematics}\label{sec_Kinematics}

We denote the initial and final particle momenta as follows:
\begin{equation}\label{proc_BH}
p (p_1) + e^{\pm} (k_1) \to p (p_2) + e^{\pm} (k_2) + \gamma (k) \left[ + \gamma (k') \right].
\end{equation}
The energies of the final(initial) electron and the photon are denoted by $E_e'$($E_e$) and $E_{\gamma}$.
We consider the phase space region with $E_{\gamma} > E_{\gamma}^{cut} \sim 5$~GeV and the polar angle $\Theta_{\gamma} < 0.001$.
The electrons can be detected if $4 < E_e' < 9$~GeV.
The initial beam energies at HERA are~\cite{ZeusLumi}:
\begin{align*}
E_{p} & = 460, \: 575 \: \text{and} \: 920 \: \text{GeV}, 
\\
E_e & = 27.6 \: \text{GeV}.
\end{align*}

The momentum transferred by the proton is denoted by $q$:
\begin{equation*}
q=p_2-p_1, \quad Q^2=-q^2>0.
\end{equation*}
The variation of $Q^2$ is very large:
\begin{equation} \label{Q2limBH}
Q^2_{min} \approx \frac{m_e^4 m_p^2}{s^2 (E_{e}\slash E_{\gamma}^{cut}-1)^2 } \approx 3 \cdot 10^{-25} \: \text{GeV}^2,
\qquad
Q^2_{max} = \lambda (s,m_p^2,m_e^2) \slash s \approx 10^5 \: \text{GeV}^2.
\end{equation}

The process of the lepton pair creation is also considered ($l=e,\mu$):
\begin{equation}\label{proc_Peee}
p (p_1) + e^{\pm} (k_1) \to p (p_2) + e^{\pm} (k_2) + l^{+} (k_3) + l^{-} (k_4).
\end{equation}
It affects the electron tagger only and is expected to be a background for the Bethe-Heitler process.

The range of $Q^2$ for this process is:
\begin{equation}
\begin{split}
{\left. Q^2_{min} \right|}_{l=e} & \approx \frac{64 m_e^4 m_p^2}{s^2} \sim 4 \cdot 10^{-22} \: \text{GeV}^2,
\qquad
{\left. Q^2_{min} \right|}_{l=\mu}
\approx \frac{36 m_{\mu}^4 m_p^2}{s^2} \sim 3 \cdot 10^{-13} \: \text{GeV}^2,
\\
{\left. Q^2_{max} \right|}_{l=e,\mu} & = \frac{1}{2 s} \left( \lambda- \delta_m (s+m_p^2-m_e^2)
+ \lambda^{1/2} \sqrt{  \lambda- 2 \delta_m (s+m_p^2-m_e^2) + \delta_m^2}
\right)
\approx \frac{\lambda}{s}
\sim 10^5 \: \text{GeV}^2
\end{split}
\end{equation}
with $\lambda = \lambda (s,m_p^2,m_e^2) = 4 \CplPow{p_1}{k_1}{2} - 4 m_e^2 m_p^2$ and $\delta_m = \left( m_e + 2 m_l \right)^2 - m_e^2$.

\begin{figure}[htb]
\centering
\includegraphics[width=.49\linewidth, angle=0]{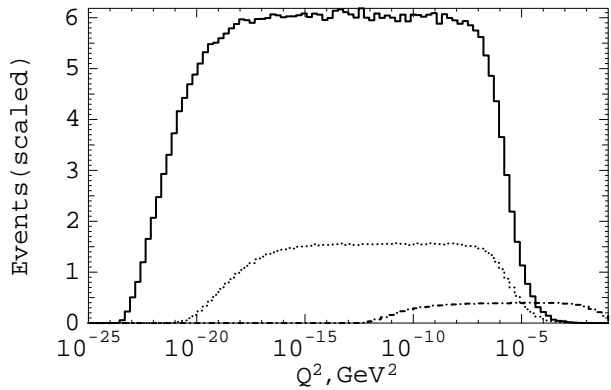}
\caption{
The $Q^2$-distribution
for the processes; $p e^{\pm} \to p e^{\pm} \gamma$ (solid line, no cuts on $E_e'$),
$p e^{\pm} \to p e^{\pm} e^{+} e^{-}$ (dotted line, $4<E_e'<9$~GeV, scaled by factor $\times 10^2$),
$p e^{\pm} \to p e^{\pm} {\mu}^{+} {\mu}^{-}$ (dashed-dotted line, $4<E_e'<9$~GeV, scaled by factor $\times 10^6$).
}
\label{fig_Q2}
\end{figure}

The actual distributions of the $Q^2$ values for the considered processes are shown in Fig.~\ref{fig_Q2}.
The effective $Q^2$ region is much smaller than the allowed region due to the matrix element.
In this paper, values of $Q^2>10^{-3} \: \text{GeV}^2$ are in fact neglected.

\section{Matrix element calculation} \label{sec_ME}

For the one-photon exchange process the cross section is written in terms of the leptonic and hadronic tensors:

\begin{equation*}
d \sigma = \frac{1}{I} \sum_{spins} {|M|}^2 d \Gamma = \frac{1}{2 \lambda^{1/2}(s,m_e^2,m_p^2)} \frac{1}{Q^4} \sum_{spins} L_{\mu \nu} W^{\mu \nu} d \Gamma.
\end{equation*}

The leptonic tensor is composed of the electromagnetic current vectors as:

\begin{equation*}
L_{\mu \nu}=\sum_{spins} j_{\mu} j^{*}_{\nu} = \sum_{spins}
\bar{u}_{k_2} e \hat{\Gamma}_{\mu} [\lambda_2,\lambda,...] u_{k_1}
\bar{u}_{k_1} e \hat{\Gamma}_{\nu}^{+} [\lambda_2,\lambda,...] u_{k_2}.
\end{equation*}
The plane wave approximation is used in spite of
the macroscopic interaction distance at the lowest edge of the $Q^2$ region ($r \sim 0.1$ mm at $Q^2 \sim 10^{-24} \text{GeV}^2$).

The hadronic tensor for unpolarized protons is written as:
\begin{equation*}
W_{\mu \nu}=-e^2 F_1 g_{\mu \nu} + e^2 F_2 \frac{{p_1}_{\mu}{p_1}_{\nu}}{m_p^2}, \quad
F_1=Q^2 G_M^2, \quad
F_2=4 m_p^2 \frac{G_E^2 + \tau G_M^2}{1+\tau}.
\end{equation*}
Here $G_E$ and $G_M$ are the Sachs form factors and $\tau=Q^2/4 m_p^2$.
We use the dipole parametrization:
\begin{equation} \label{ff_Hof}
G_E = \frac{1}{{\left( 1+Q^2/{\Lambda_D^2} \right)}^2}, \quad G_M = \mu_p G_E.
\end{equation}
Here $\Lambda_D^2=0.71 \: \text{GeV}^2$ and $\mu_p$ is the magnetic moment of proton.
The precision of the dipole formula is very poor for $Q^2$ values greater than $0.1 \: \text{GeV}^2$~\cite{HydeWrightDeJager05}.
We use it here only to check whether the proton size effects are visible in the considered processes.
As can be seen from Fig.~\ref{fig_Q2},
the effective region of the transferred momentum is $Q^2 < 10^{-5} \: \text{GeV}^2$ for the
photon-
and $e^+ e^-$-pair production processes.
The dipole form factor~(\ref{ff_Hof}) differs from unity by $0.003 \%$ in that region.
On the other hand, modern form factor formulae~\cite{HydeWrightDeJager05,FriedrichWalcher03}
were mostly fitted with $Q^2 > 0.1 \text{GeV}^2$ experimental data.
They have a similar $(1-G_E)$-value at very low $Q^2$.
Hence we may either use the simple dipole formula~(\ref{ff_Hof}) or neglect proton structure effects accepting the value of $(1-G_E(Q^2_{max})) \sim 0.003 \%$ as the systematic error of the calculation.
The error of our estimate for the $pe \to pe {\mu}^+ {\mu}^-$ process ($10^{-12} < Q^2 < 1 \: \text{GeV}^2$) is therefore approximately $20 \%$
due to the form factor uncertainty.

The symbolic algebra program ALHEP~\cite{alhep} is used for analytical computations.
It creates diagrams using the Standard Model Lagrangian, calculates matrix elements and reduces tensor integrals over the virtual particles' phase space to scalar ones.
Finally the matrix element is written in terms of the scalar products of the initial and final momentum vectors.
The infrared divergences are regularized using a fixed photon mass parameter.
The contributions of the soft radiative corrections are integrated analytically over the photon energy
in the region $m_{\gamma}^{IR} <E_{\gamma}<E^{SH}_{\gamma}$ in the laboratory frame.
The basics are well described in Ref.~\cite{Denner93}.
The $E^{SH}_{\gamma}$ energy cutoff separates the soft and hard bremsstrahlung contributions.
We vary it in the range of $10^{-8}-10^{-4}$~GeV to control the generator consistency.
The \verb|LoopTools|~\cite{LoopTools} package is used for the numerical estimation of loop integrals.
It was recompiled with \verb|real*16| precision to avoid internal instability at very small $Q^2$.

The matrix elements suffer from huge numerical cancellations in the considered phase space region.
The first problem appears when calculating momentum couplings by contracting the corresponding vectors.
It is solved as described in Sec.~\ref{sec_Generator}.
Another problem has to do with the cancellation of peaking terms in the matrix element.
Let us consider typical values of the scalar products for the Born process:
\begin{align*}
Q^2 \sim \Cpl{p_1}{q} \sim \Cpl{p_2}{q} & \sim 10^{-22} \ldots 10^{-5} \: \text{GeV}^2, \\
| \Cpl{k}{q} | \sim | \Cpl{k_1}{q} | \sim | \Cpl{k_2}{q} | & \sim 10^{-6} \: \text{GeV}^2, \\
\Cpl{k_1}{k_2} \sim \Cpl{k}{k_1} \sim \Cpl{k}{k_2} & \sim 10^{-6} \: \text{GeV}^2, \\
\Cpl{p_i}{k_1} \sim \Cpl{p_i}{k_2} \sim \Cpl{p_i}{k} & \sim 10^{4} \ldots 10^{5} \: \text{GeV}^2.
\end{align*}
The major cancellation appears due to $\Cpl{p_i}{k_j}$ couplings.
The $\Cpl{p_2}{k_j}$ and $\Cpl{p_1}{k}$ values are easily expressed via the others.
But the two remaining couplings ($\Cpl{p_1}{k_1}$ and $\Cpl{p_1}{k_2}$) are independent and may cancel numerically.

We use the substitution:
\begin{equation}
\delta_{1}
= \Cpl{k}{k_1} \Cpl{p_1}{k_1} - \Cpl{k}{k_2} \Cpl{p_1}{k_2}.
\end{equation}
Typical values for $\delta_{1}$ are:
\begin{equation} \nonumber
\delta_{1} \sim 10^{-1}-10^{-1} \approx 10^{-9} \: \text{GeV}^4.
\end{equation}
Thus calculating $\delta_{1}$ precisely as 
$\delta_{1} = \Cpl{(\Cpl{k}{k_1}{k_1}_{\alpha}-\Cpl{k}{k_2}{k_2}_{\alpha})}{p_1^{\alpha}}$
one can avoid one of the two $\Cpl{p_1}{k_i}$ couplings.
In this way the Born differential cross section can be written as:
\begin{equation} \label{cs_Born}
\begin{split}
\frac{d \sigma}{d \Gamma} = {}& \frac{e^6}{q^4} \frac{A_{1} F_1({Q^2}) - A_{2} F_2({Q^2})}{2 \lambda^{1/2}(s,m_e^2,m_p^2)}, \\
A_{1} ={}&
\left( \frac{4\Cpl{k}{k_2}}{\Cpl{k}{k_1}}+\frac{2 m^2_e (2 m^2_e + q^2)}{\CplPow{k}{k_1}{2}}\right)
\left(1+\frac{\CplPow{k}{k_1}{2}}{\CplPow{k}{k_2}{2}}\right)
-
\frac{ 4 \Cpl{k_1}{k_2} (2 m^2_e+q^2) }{ \Cpl{k}{k_1} \Cpl{k}{k_2} },
\\
A_{2} = {}&
\frac{4 m^2_e \delta_{1}^2}{\CplPow{k}{k_1}{2} \CplPow{k}{k_2}{2} m^2_p}
+ \frac{2 \Cpl{k}{k_2}}{\Cpl{k}{k_1}} + \frac{2 \Cpl{k}{k_1}}{\Cpl{k}{k_2}}
+
\frac{q^2}{m^2_p}
\left(
\left( \frac{2 \CplPow{k_1}{p_1}{2}}{\Cpl{k}{k_1} \Cpl{k}{k_2}} - \frac{2 \Cpl{k_1}{p_1}}{\Cpl{k}{k_1}} + \frac{m^2_e m^2_p}{\CplPow{k}{k_1}{2}} \right)
\left( 1 + \frac{\CplPow{k}{k_1}{2}}{\CplPow{k}{k_2}{2}} \right)
-
\right.
\\
&
- \frac{4 \Cpl{k_1}{p_1} \delta_{1}}{\CplPow{k}{k_2}{3}}
- \frac{q^2 \Cpl{k_1}{p_1}}{\CplPow{k}{k_2}{2}}
+ \frac{q^2 \Cpl{k_1}{p_1}}{\Cpl{k}{k_1} \Cpl{k}{k_2}}
+ \frac{2 \delta^2_{1}}{\Cpl{k}{k_1} \CplPow{k}{k_2}{3}} +
\\
&
\left.
+ \frac{2 \delta_{1} m^2_e}{\CplPow{k}{k_1}{2} \Cpl{k}{k_2}}
+ \frac{2 \delta_{1} m^2_e}{\Cpl{k}{k_1} \CplPow{k}{k_2}{2}}
+ \frac{2 \delta_{1}}{\CplPow{k}{k_2}{2}}
+ \frac{\delta_{1} q^2}{\Cpl{k}{k_1} \CplPow{k}{k_2}{2}}
- \frac{2\Cpl{k_1}{k_2} m^2_p}{\Cpl{k}{k_1} \Cpl{k}{k_2}}
+ \frac{q^2 m^2_e}{\Cpl{k}{k_1} \Cpl{k}{k_2}}
\right)
.
\end{split}
\end{equation}
Here the remaining $\Cpl{k_1}{p_1}$-terms appear coupled to the tiny $Q^2$ factor.

The numerical behavior of this formula is stable enough to be used with fast 8-byte variables in the analysis.
Alternatively this matrix element can be written in a more compact but still numerically stable way
in terms of Levi-Civita tensor couplings~\cite{BH_Gaemers88}.

For the higher-order processes there is no simple substitution available,
and the straightforward expansion in terms of one major term is not possible.
At least $4$ independent couplings may occur as major terms in various phase space regions.
This problem can be partially solved using the Levi-Civita tensor couplings method~\cite{BH_Gaemers88}.
The other possible solution is to split the phase space and to adjust the matrix element separately for every part.
This method leads to enormous growth of the code and problems during the debugging step.
In this analysis we solve the precision problem using high-precision floating point numbers~\cite{QD}.
The loop and the hard bremsstrahlung corrections~(\ref{proc_BH}) and the lepton pair creation process~(\ref{proc_Peee})
are calculated with 16-byte variables.
They allow to use the same matrix element form in the entire phase space region.

For cross-check purposes several matrix element representations (using various sets of independent variables) are compared.
The results are similar.
However, numerical precision for different representations is quite different.
Expressions which depend on the transferred momentum $\Cpl{q}{k_i}$-couplings are much more efficient.
Some representations require at least 32-byte variables for proper evaluation.
The splitting parameter and renormalization constants are also varied to check the consistency of the generator.

The higher-order matrix elements are not compact enough to be given in this paper.
One can find all the formulae (converted to \verb|Mathematica| format)
together with \verb|ALHEP|-scripts used for their derivation in Ref.~\cite{bhgen_doc}.

\section{Generator} \label{sec_Generator}

The typical adaptive Monte-Carlo generator is based on iterative phase space splitting
according to the distribution of the integrated function (or its derivative).
At the first (integration) step, the appropriate splitting grid is created and at the second (generation) step, the events are sampled.
However, the narrow peaks of the function may lead to significant integration errors.
Since only a finite number of events is generated in every cell, the peak may remain undetected.
The integration in the forward region is complicated by the fact
that the effective region of the phase space is negligible as compared to the total range of the variables.
For example, the \verb|FOAM|~\cite{foam} generator (with default parameters) being efficient for other purposes
failed to integrate the Born cross section~(\ref{cs_Born}) in the whole phase space region~(\ref{Q2limBH}).

Cuts may also lead to integration errors.
In our case several cuts must be applied at the integration step:
\begin{itemize}
\item Minimum energy of the detected photon, $E^{min}_{\gamma} \sim 5$~GeV;
\item Minimum energy of the undetected photon, $E^{SH}_{\gamma} \sim 10^{-8} \cdots 10^{-5}$~GeV (soft and hard bremsstrahlung separator);
\item Energy cut on the detected electron in the $l^+l^-$-production process: $4 < E_e < 9$~GeV.
The cross section with and without this cut is about $0.1$ and $4.0$~mb respectively.
This cut needs to be applied at the integration level.
Otherwise the event generation is slowed by a factor $40$.
\end{itemize}

Therefore a new adaptive Monte-Carlo integration program was created.
It uses the photon energy as integration variable, thus avoiding $E_{\gamma}$ cuts.
The main idea is to carefully tune the sub-space splitting algorithm
to minimize the integration time and the amount of memory used.
The distribution of the derivative is examined to detect narrow peaks.
A separate splitting algorithm is used if a function cut is detected inside the cell.
Generated function values are re-used for further sub-cell integration wherever possible.
The maximum number of steps of the phase space splitting is about $110$ for $2 \to 3$ processes and $130$ for $2 \to 4$ processes.
The statistical error of the integration is distributed according to the function value: $\delta I_{i} \propto I_i$.
For cross-check purposes, the Born process is partially integrated with the {\tt FOAM}~\cite{foam} generator.
The {\tt FOAM} results are stable in the sub region $Q^2 > 10^{-9} \: \text{GeV}^2$ and are in good agreement with our results.


Another problem is to compose the appropriate phase space reconstruction procedure.
This procedure is used to map a hypercube point $r_i \in (0,1)^N$ produced by the generator to the phase space of final particles.
The standard method of factorizing the phase space to 2-body decay sub spaces \cite{Weinzierl} is used:
\begin{equation}\nonumber
d {\Gamma_n} (p_n) = (2 \pi)^{-1} \, d{p^2_{n-1}} \, d {\Gamma_2 (p_{n} \to k_n + p_{n-1}) } \, d {\Gamma_{n-1}} (p_{n-1}).
\end{equation}
However, the vectors of the 2-particle sub space are not calculated in the rest frame of the total momentum~$p_n$
since Lorentz-transformations in the forward region lead to about $6$-digit precision loss.
The procedure is described in the following:

We generate the recoiling proton first:
\begin{equation} \label{dG_p2}
\int d \Gamma_{n} =
\frac{1}{2 \pi \lambda^{1/2} (s,m_p^2,m_e^2) }
\int\limits_0^{2 \pi} d {\phi_{z}}
\int
Q^4 d Q^{-2}
\int
d {p_{n-1}^2}
\int d \Gamma_{n-1}.
\end{equation}
Here $\phi_{z}$ is the free axial rotation angle for the event,
$p_{n-1}$ and $\Gamma_{n-1}$ are the momentum and the phase space of the remaining particles
(e.g. $p_{n-1}=k+k_2$ for the Born process),
and the term $Q^4$ cancels one in the denominator of the matrix element.
The limits for $Q^2$ are defined in Sec.~\ref{sec_Kinematics} for each process.
The limits on $p_{n-1}^2$ are:
\begin{equation}\label{limits_MqQ}
{p_{n-1}^2} {\mid}_{min} = (\sum\limits_{n-1}{m_i})^2, \quad {p_{n-1}^2} {\mid}_{max} = m_e^2 - \frac{Q^2 (s+m_p^2-m_e^2)}{2 m_p^2}
+ \frac{\lambda^{1/2}(s,m_p^2,m_e^2)}{2 m_p^2} \sqrt{Q^4 +4 m_p^2 Q^2}.
\end{equation}

For the $pe \to pe \gamma$ process the most convenient order of the remaining integration is:
\begin{equation}
\int d \Gamma_{n-1} =
\int\limits_0^{2 \pi}
\int\limits_{E_{\gamma}^{min}}^{E_{\gamma}^{max}}
\frac{
d {\phi_{\vec{k} + \vec{k_2}}} \: d E_{\gamma}
}{(4 \pi)^2 |\vec{k}+\vec{k_2}|}
.
\end{equation}
Here $\phi_{\vec{k}+\vec{k_2}}$ is the angle of the plane spanned by the electron and the photon
with respect to the direction of their sum vector $\vec{k} + \vec{k_2}$ in the lab. frame.
The integration limits on the energy are defined by the general formulae:
\begin{equation}
\label{limits_E}
\begin{split}
E_{i}^{min}
& =
\max \left\{
\frac{1}{2 p^2} \left( p^0 (p^2 - (p-k_i)^2 + m_i^2) - |\vec{p}| \lambda^{1/2}(p^2,(p-k_i)^2,m_i^2) \right)
, \: E^{cut}_{min}
\right\},
\\
\quad
E_{i}^{max}
& =
\min \left\{
\frac{1}{2 p^2} \left( p^0 (p^2 - (p-k_i)^2 + m_i^2) + |\vec{p}| \lambda^{1/2}(p^2,(p-k_i)^2,m_i^2) \right)
, \: E^{cut}_{max}
\right\}.
\end{split}
\end{equation}
For the Born process here: $p=p_{n-1}=k+k_2$, $k_i=k_{\gamma}$, $m_i=0$, $(p-k_i)^2=m_e^2$ and the $E^{cut}_{max}$ is unused.
Finally, the limits on the photon energy are:
\begin{equation} \label{limits_Eg}
E_{\gamma}^{min}
=
\max \left\{
\frac{p_{n-1}^2 - m_e^2}{2 p_{n-1}^2} (p_{n-1}^0-|\vec{p}_{n-1}|),
\: E^{cut}_{\gamma}
\right\},
\quad
E_{\gamma}^{max}
=
\frac{p_{n-1}^2 - m_e^2}{2 p_{n-1}^2} (p_{n-1}^0+|\vec{p}_{n-1}|).
\end{equation}

For the two photon bremsstrahlung process $p_{n-1} = k + k_2 + k'$ and the phase space integral is given by:
\begin{equation}
\int d \Gamma_{n-1} =
\int\limits_{m_e^2}^{(\sqrt{p^2_{n-1}}-m_e)^2}
\frac{d {(k+k_2)^2}}{2 \pi}
\int\limits_0^{2 \pi}
\int\limits_{E^{SH}_{\gamma}}^{E_{\gamma'}^{max}}
\frac{
d {\phi_{\vec{k}+\vec{k_2}+\vec{k'}}} \: d {E_{\gamma}'}
}{(4 \pi)^2 |\vec{k}+\vec{k_2}+\vec{k'}|}
\int\limits_0^{2 \pi}
\int\limits_{E_{\gamma}^{min}}^{E_{\gamma}^{max}}
\frac{
d {\phi_{\vec{k}+\vec{k_2}}} \: d {E_{\gamma}}
}{(4 \pi)^2 |\vec{k}+\vec{k_2}|}
.
\end{equation}
Here the limits on $E_{\gamma}$ are defined by Eq.~(\ref{limits_Eg}) with $p_{n-1}=k+k_2$
and the additional condition $E_{\gamma} \ge E_{\gamma}'$.
The $E_{\gamma}'$-limits are obtained from Eq.~(\ref{limits_Eg}) by the substitutions
$m_e^2 \to (k+k_2)^2$ and $E^{cut}_{\gamma} \to E^{SH}_{\gamma}$ (the soft and hard bremsstrahlung separator).

Similar formulae are used for the lepton pair production process ($p_{n-1}=k_2+k_3+k_4$):
\begin{equation}
\int d \Gamma_{n-1} =
\int\limits_{4 m_l^2}^{(\sqrt{p^2_{n-1}}-2 m_l)^2} \frac{d {(k_3+k_4)^2}}{2 \pi}
\int\limits_0^{2 \pi}
\int\limits_{E_{e'}^{min}}^{E_{e'}^{max}}
\frac{
d {\phi_{\vec{k_2}+\vec{k_3}+\vec{k_4}}} \: d {E_{e}'}
}{(4 \pi)^2 |\vec{k_2}+\vec{k_3}+\vec{k_4}|}
\int\limits_0^{2 \pi}
\int\limits_{E_{l^{-}}^{min}}^{E_{l^{-}}^{max}}
\frac{
d {\phi_{\vec{k_3}+\vec{k_4}}} \: d {E_{l^{-}}}
}{(4 \pi)^2 |\vec{k_3}+\vec{k_4}|}
.
\end{equation}
Here the limits on $E_{e}'$ are given by Eq.~(\ref{limits_E}) with
$p=p_{n-1}$, $m_i=m_e$, $(p-k_i)^2=(k_3+k_4)^2$ and the cuts on the final electron are the $E^{cut}_{min/max}$ parameters.
The integration limits for the second lepton energy $E_{l^{-}}$ are calculated using the same formula with
$p=k_3+k_4$, $m_i=m_l$, $(p-k_i)^2=m_l^2$ and no additional cuts for the $\mu^{+} \mu^{-}$-production process.
For the case $l=e$ we set $E^{max}_{l^{-}} = E_{e}'$ in Eq.~(\ref{limits_E}).
Alternatively, one may apply the detector cuts on the second electron together with the factor $1/2$ to avoid double-counting of identical particles.

Calculating the scalar products by contracting previously reconstructed momenta also leads to large numerical errors.
Therefore we define products in the generator using the original $r_i$-values.

The new code saves up to $30$ decimal digits of precision
when compared to the typical algorithm that uses momenta in the two-body rest frames~\cite{Weinzierl}.
The Born process converges with the standard 8-byte floating point variables.
Sixteen-byte floating point variables (\verb|QD| package~\cite{QD}) are used for the higher-order process calculation.

A different set of integration variables is implemented to cross check the generator.
The recoil proton is still reconstructed according to Eqs.~(\ref{dG_p2}-\ref{limits_MqQ})
and the sequential approach from Ref.~\cite{Weinzierl} is used for the remaining particles.
Here the energy cuts are applied on the integrating function
and the center-of-mass frames are used for momentum reconstruction.
The result of the integration agrees with the previous one
but the generator requires $32$-byte variables (and huge computer resources) for numerical stability.

\begin{figure}[ht]
\centering
\includegraphics[height=3.5in, angle=0]{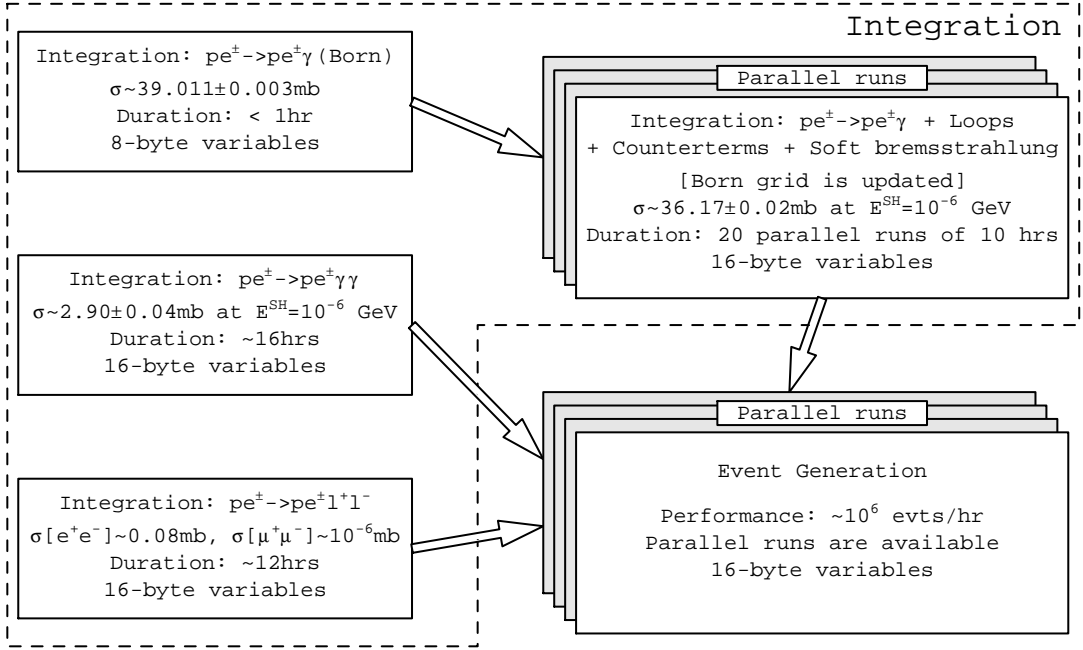}
\caption{
The generator program structure.
The cross section values correspond to $E_{\gamma}^{cut}=10$~GeV for the photoproduction processes
and $4 < E_e < 9$~GeV for the lepton pair production process.
}
\label{fig_Gen}
\end{figure}

Due to the use of high-precision floating point numbers, the performance of the generator must be carefully tuned.
Both integration and generation steps require sufficient computer resources: up to 2Gb of memory and about a day of the CPU time.
To optimize the generator performance we make parallel runs wherever possible.
Thus the integration of the sub processes is performed separately
and events are generated in parallel runs using independent sets of random numbers.
We also integrate the loop- and factorisable soft bremsstrahlung corrections
using the previously created Born-process pattern.
At every point in phase space the Born matrix element is within $10-30\%$ of the corrected one,
$|M_{Born}|^2 \approx |M_{Born}|^2 + 2 \Re (M_{Born}^{*} M_{loop}) + |M^{R}_{soft}|^2$.
Therefore we use the Born process integration grid and simply update the integral value in every sub cell
(using the distributed computation again).
The scheme of optimal usage of the generator is outlined in Fig.~\ref{fig_Gen}.

\section{Results} \label{sec_Results}

We consider the process $p e^{\pm} \to p e^{\pm} \gamma$ with at least one high energy photon in the final state ($E>E^{cut}_{\gamma}$).
The value of $E^{cut}_{\gamma}$ is set to $10$~GeV.
We also simulate the energy acceptance of the photon spectrometer~\cite{ZeusLumi} using an approximate formula.
For the simulation of the lepton-photon coincidence, an additional cut on the lepton $4 < E_{e} < 9$~GeV is used.
The background lepton pair creation process~(\ref{proc_Peee}) is also considered.
When the two identical particles ($e$ or $\gamma$) appear in the detectable region only the hardest one is observed.
The PDG'09
values are used for physical constants.
The results for the cross sections are given in Table~\ref{tbl_CS} for different beam energies.
All other results are for $E_p=920$~GeV.
\begin{table}[htb]
\centering
\begin{tabular}{|c|c|c|c|c|}
\hline
Cuts & Process & $\sigma(460 \: \text{GeV}), \: \text{mb}$ & $\sigma(575 \: \text{GeV}), \: \text{mb}$ & $\sigma(920 \: \text{GeV}), \: \text{mb}$
\\[3pt]
\hline
$E_{\gamma} > 10$~GeV
&
\begin{tabular}{@{}c@{}} Bethe-Heitler formula~(\ref{classic_E}) \\ $p e^{\pm} \to p e^{\pm} \gamma$ (Born) \\ $p e^{\pm} \to p e^{\pm} \gamma$ (+ RC) \end{tabular}
&
\begin{tabular}{@{}c@{}} $37.505$ \\ $37.504$ \\ $37.58$ \end{tabular}
&
\begin{tabular}{@{}c@{}} $37.991$ \\ $37.990$ \\ $38.06$ \end{tabular}
&
\begin{tabular}{@{}c@{}} $39.012$ \\ $39.011$ \\ $39.07$ \end{tabular}
\\ \hline
\begin{tabular}{@{}c@{}} $E_{\gamma} > 10$~GeV \\ $4 < E_{e} < 9$~GeV \end{tabular}
&
\begin{tabular}{@{}c@{}} Bethe-Heitler formula~(\ref{classic_E}) \\ $p e^{\pm} \to p e^{\pm} \gamma$ (Born)
\\ $p e^{\pm} \to p e^{\pm} \gamma$ (+ RC) \\ $p e^{\pm} \to p e^{\pm} e^{+} e^{-}$ \\ $p e^{\pm} \to p e^{\pm} {\mu}^{+} {\mu}^{-}$ \end{tabular}
&
\begin{tabular}{@{}c@{}} $8.35$ \\ $8.35$ \\ $8.41$ \\ $0.078$ \\ $\sim 1.1 \cdot 10^{-6}$ \end{tabular}
&
\begin{tabular}{@{}c@{}} $8.46$ \\ $8.46$ \\ $8.53$ \\ $0.079$ \\ $\sim 1.1 \cdot 10^{-6}$ \end{tabular}
&
\begin{tabular}{@{}c@{}} $8.69$ \\ $8.69$ \\ $8.81$ \\ $0.082$ \\ $\sim 1.2 \cdot 10^{-6}$ \end{tabular}
\\ \hline
\end{tabular}
\caption{
Cross sections for the processes considered here.
The precision of the ${\mu}^+ {\mu}^-$-cross section is significantly suppressed by the proton form factor uncertainty at $Q^2 \sim 1 \: \text{GeV}^2$ (see Sec.~\ref{sec_ME}).
}
\label{tbl_CS}
\end{table}

The uncertainty of the the Born process integration is less than $0.01 \%$.
The generator error for the process with the one-loop radiative correction (RC) is about $0.2 \%$.
This uncertainty is dominated by the precision of the sub process $pe \to pe \gamma \gamma$
in the neighborhood of the soft and hard bremsstrahlung separation border $E_{\gamma'} \sim E^{SH}_{\gamma}$.
The large peak of the radiative process here results in the systematic error of $\sim 0.1-0.2 \%$ in the total cross section.
Better precision can be achieved by increasing the integration time and computer resources used.
However, since the total error of the ZEUS luminosity measurement is about $1-2 \%$
the theoretical error of about $0.2 \%$ is acceptable for the cross section calculation.
The uncertainty for $e^{+}e^{-}$-pair production is about $0.1 \%$ of the total electron tagger signal.

\begin{figure}[htb]
\leavevmode
\begin{minipage}[h]{.49\linewidth}
\centering
\includegraphics[width=\linewidth, angle=0]{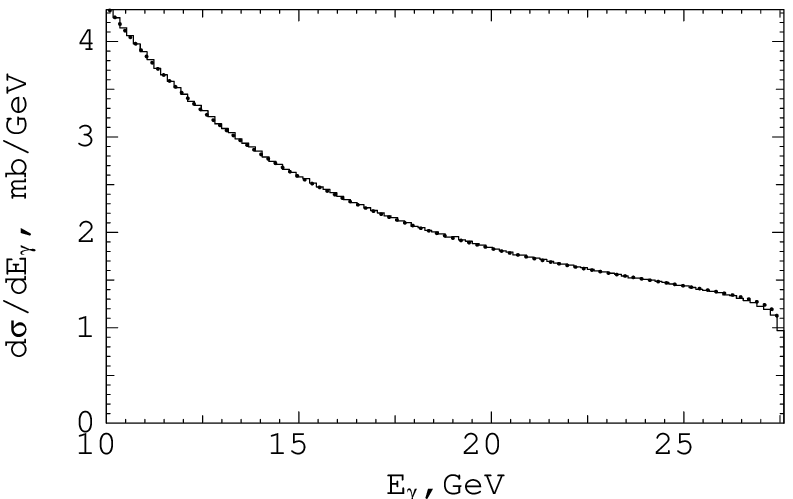}
\end{minipage}
\begin{minipage}[h]{.49\linewidth}
\centering
\includegraphics[width=\linewidth, angle=0]{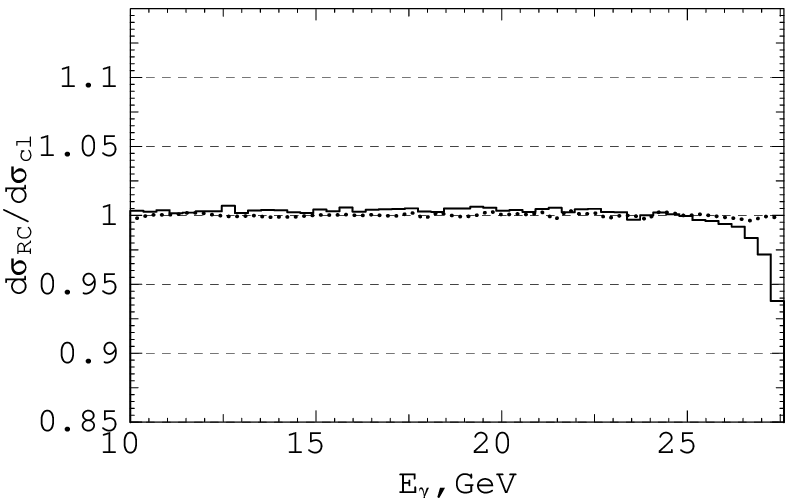}
\end{minipage}
\caption{
The photon energy spectrum (left) and its ratio to the Bethe-Heitler formula~(\ref{classic_E}) prediction (right).
The dashed line on the left plot is the shape of the classical distribution.
The ratio of the Born process cross section to the Bethe-Heitler formula is also presented (dotted line).
The discrepancy of the total cross section is within $0.2 \%$ value.
}
\label{fig_Eg}
\end{figure}

\begin{figure}[htb]
\centering
\leavevmode
\begin{minipage}[h]{.49\linewidth}
\centering
\includegraphics[width=\linewidth, angle=0]{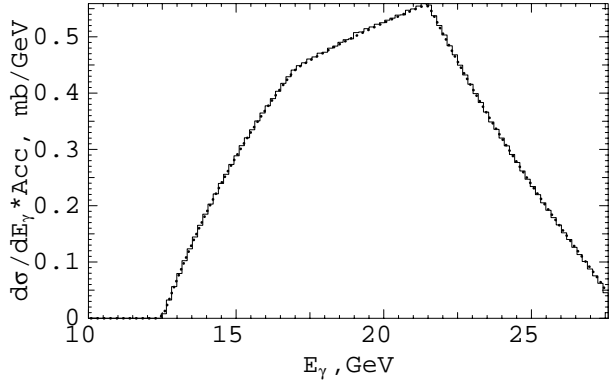}
\end{minipage}
\caption{
The photon energy spectrum (solid line) and Bethe-Heitler formula values (dotted line) after the spectrometer acceptance is applied.
The difference is not visible. The total cross sections agree within $0.2\%$.
}
\label{fig_EgAcc}
\end{figure}

The Born cross section and the photon energy spectrum coincide with the classical Bethe-Heitler formula~(\ref{classic_E}) predictions.
When the radiative corrections are applied, the total cross section is still equal to the classical one within the integration error.
But the energy spectrum is slightly different at very high $E_{\gamma}$ values (Fig.~\ref{fig_Eg}).
This is exactly the region of the high systematic error of the radiative sub-process integration.
However, this region is suppressed by the small acceptance of the photon spectrometer~\cite{ZeusLumi}.
Therefore there is no significant difference in the energy spectrum after the detector acceptance is applied (Fig.~\ref{fig_EgAcc}).

\begin{figure}[htb]
\centering
\leavevmode
\begin{minipage}[h]{.49\linewidth}
\centering
\includegraphics[width=\linewidth, angle=0]{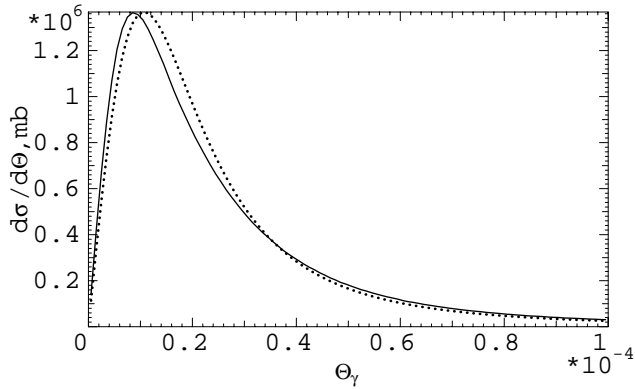}
\end{minipage}
\caption{
The photon angular spectrum (solid line)
and the classical formula~(\ref{classic_Theta}) values (dashed line).
}
\label{fig_Theta}
\end{figure}

The angular spectrum of the photons differs from the expected classical shape~(\ref{classic_Theta}) (see Fig.~\ref{fig_Theta}).
However, the divergence of the electron beam in ZEUS ($\sim 0.09-0.21$~mrad~\cite{ZeusLumi})
is much greater than the mean photon emission angle.
Thus the actual angular spectrum shape should not affect the luminosity measurement.
To estimate the expected effect we consider the size of the photon detector.
We estimate the fraction of events outside the spectrometer depending on the angular distribution used.
The exit window of the ZEUS spectrometer system has a diameter of about $10$~cm and is located at $92.5$~m from the interaction point~\cite{ZeusLumi}.
To simulate the angular spread of the incoming electron beam
we used a Gaussian distribution with $\sigma_x=21 \cdot 10^{-4}, \sigma_y=9 \cdot 10^{-4}$.
The results, ignoring the angular acceptance of the spectrometer, are given in Table~\ref{tbl_Ang}.
The effect of the photon angular spectrum is less than $0.05\%$.
This uncertainty is below the required generator error and can be neglected.
If one ignores the scattering angle and only considers $\Theta_{\gamma}=0$ for all photons the difference is about $0.4\%$.
\begin{table}[htb]
\centering
\begin{tabular}{|c|c|c|c|}
\hline
Detector radius, mm
& $\Theta_{\gamma}$ by generator &
$\Theta_{\gamma}$ by Eq.~(\ref{classic_Theta}) &
$\Theta_{\gamma} = 0$
\\[3pt] \hline
40
& 5.07 & 4.97 & 4.46
\\ \hline
50
& 1.44 & 1.39 & 1.12
\\ \hline
60
& 0.39 & 0.36 & 0.23
\\ \hline
70
& 0.14 & 0.12 & 0.04
\\ \hline
\end{tabular}
\caption{
Fractions of events (in $\%$) outside the photon detector acceptance
depending on the angular distribution used and the size of the ZEUS spectrometer exit window.
The angular spread of initial electron beam is simulated with $\sigma_x=21 \cdot 10^{-4}$, $\sigma_y=9 \cdot 10^{-4}$.
}
\label{tbl_Ang}
\end{table}

\begin{figure}[htb]
\centering
\leavevmode
\begin{minipage}[h]{.49\linewidth}
\centering
\includegraphics[width=\linewidth, angle=0]{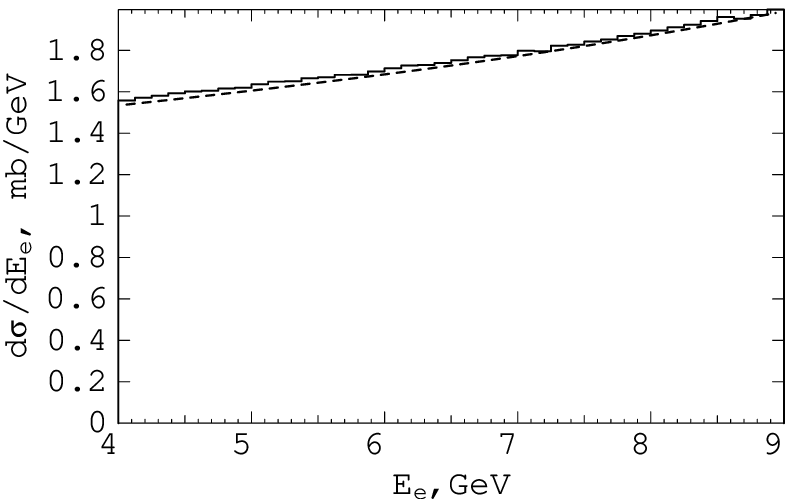}
\end{minipage}
\begin{minipage}[h]{.49\linewidth}
\centering
\includegraphics[width=\linewidth, angle=0]{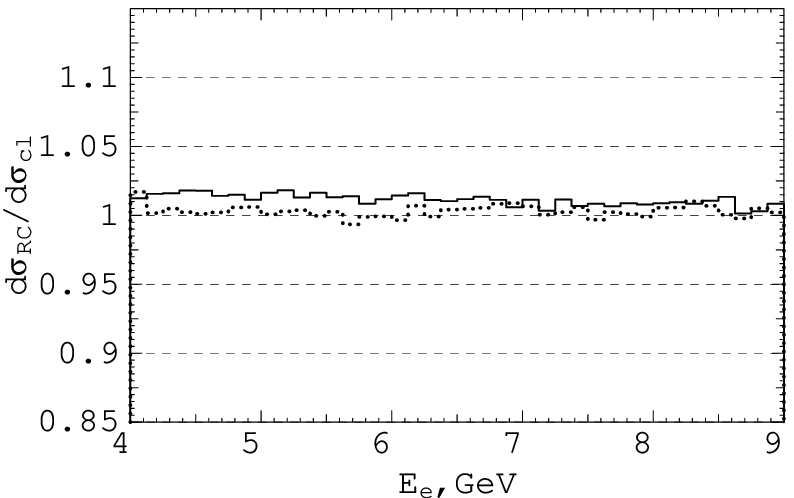}
\end{minipage}
\hfill
\caption{
The electron energy spectrum (left) and its ratio to the Bethe-Heitler formula~(\ref{classic_E}) (right).
The ratio without the $e^+ e^-$-production process is also presented (dotted line).
The dashed line on the left plot is the shape of the classical distribution.
}
\label{fig_Ee}
\end{figure}

The electron energy spectrum obtained from the generator is slightly above the classical formula (see Fig.~\ref{fig_Ee}).
This arises from the additional $e^{+}e^{-}$-pair production process~(\ref{proc_Peee}).
The difference increases in the low-$E_e$ region up to a few percent.
This contribution may affect the calibration procedure of the ZEUS photon spectrometer.

\begin{figure}[ht]
\centering
\leavevmode
\begin{minipage}[h]{.49\linewidth}
\centering
\includegraphics[width=\linewidth, angle=0]{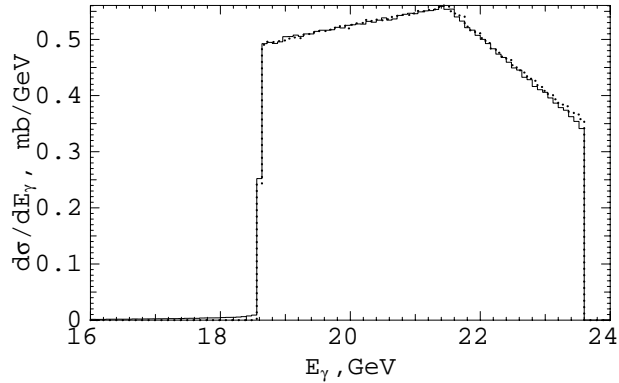}
\end{minipage}
\caption{
The photon energy spectrum (the spectrometer acceptance is applied) if the electron is detected in the region $4 < E_e < 9$~GeV.
}
\label{fig_e_g}
\end{figure}

The photon energy spectrum for $e-\gamma$ coincidence is presented in Fig.~\ref{fig_e_g}.
The deviation from the classical formula at the edges of the plot is due to
the additional momentum carried away by the undetected photon.

\section{Conclusions}

In this paper we have considered the processes that affect the luminosity measurement in the ZEUS experiment.
The basic process $p e^{\pm} \to p e^{\pm} \gamma$ has been calculated including one-loop QED radiative corrections.
The contribution of the lepton pair production process has also been estimated.

The following conclusions can be drawn:

\begin{itemize}
\item
The Born cross section and the photon energy spectrum coincide with the classical Bethe-Heitler formula values~(\ref{classic_E})
within $0.01 \%$;
\item
The angular spectrum is significantly different from the classical estimate~(\ref{classic_Theta}).
However, for the ZEUS photon spectrometer the angular effect is less than $0.05\%$;
\item
The contribution of the higher-order effects to the total cross section
of the inclusive $p e^{\pm} \to p e^{\pm} \gamma$ process is within the $0.2\%$ value;
\item
The rate of the electrons in the detector is about $1-2 \%$ above the the Bethe-Heitler formula prediction
due to the additional $e^{+}e^{-}$-pair production process.
\end{itemize}

We have achieved an error of $0.2\%$ for the cross section calculation.
This value may be treated as the theoretical uncertainty of the ZEUS luminosity measurement procedure.
Further increasing the precision is a difficult task that requires additional higher-order effects to be estimated.
However, the error of $0.2\%$ is much smaller than the systematic error of the ZEUS photon spectrometer and is acceptable for the experiment.



\begin{thebibliography}{99}

\bibitem{ZeusLumi}
M. Helbich et al.,
Nucl.Instr.Meth. {\bf A 565}(2006)572-588;
\\
J. Andruszkow et al.,
ZEUS-Note 01-004.

\bibitem{BHOrig}
H. Bethe, W. Heitler, Proc.Roy.Soc.Lond. {\bf A146}(1934)83-112.

\bibitem{BH_Gaemers88}
K.J.F. Gaemers, M. van der Horst, Nucl.Phys. {\bf B316}(1989)269-288, Erratum-ibid. {\bf B336}(1990)184.

\bibitem{BH_Horst90}
M. van der Horst, Nucl.Phys. {\bf B347}(1990)149-183.

\bibitem{Blunden03}
P.G. Blunden et al., Phys.Rev.Lett. {\bf 91}(2003)142304.

\bibitem{BH_Gaemers89}
K.J.F. Gaemers, M. Van Der Horst, J.A.M. Vermaseren, Z.Phys. {\bf C45}(1989)123-127.

\bibitem{BH_Horst90b}
M. van der Horst, Phys.Lett. {\bf B244}(1990)107-112.

\bibitem{BH_GandH}
R.L. Glueckstern, M.H. Hull, Phys.Rev. {\bf 90}(1953)1030-1035.

\bibitem{hbgen_program}
URL: \verb|http://desy.de/~vmakar/bh/|.

\bibitem{HydeWrightDeJager05} C.E. Hyde-Wright, K. de Jager, Ann.Rev.Nucl.Part.Sci. {\bf 54}(2004)217-267.

\bibitem{FriedrichWalcher03} J. Friedrich, T. Walcher, Eur.Phys.J. {\bf A17}(2003)607-623.

\bibitem{alhep} V. Makarenko, hep-ph/0704.1839; \verb|http://www.hep.by/alhep|.

\bibitem{Denner93} A. Denner, Fortsch.Phys. {\bf 41}(1993)307-420.

\bibitem{LoopTools}
T. Hahn, Nucl.Phys.Proc.Suppl. {\bf 89}(2000)231-236.

\bibitem{QD} Y. Hida, Xi.S. Li, D.H. Bailey, Tech. Report LBNL-46996(2000).

\bibitem{bhgen_doc} URL: \verb|http://desy.de/~vmakar/bh/doc|
.

\bibitem{foam} S. Jadach, Comput.Phys.Commun. {\bf 152}(2003)55-100.

\bibitem{Weinzierl} S. Weinzierl, NIKHEF-00-012.



\end{thebibliography}


\end{document}